\documentclass[journal]{IEEEtran}

\ifCLASSINFOpdf
\else
\fi

\usepackage{hyperref}        
\usepackage{url}             
\usepackage{booktabs} 
\usepackage{amsfonts} 
\usepackage{nicefrac}
\usepackage{microtype}       
\usepackage{bookmark}
\usepackage{array}
\usepackage{graphicx}
\usepackage{amsmath}
\usepackage{amsthm}
\usepackage{amsfonts} 
\usepackage{amssymb}
\usepackage{array}
\usepackage{algorithmic}
\usepackage{algorithm}
\usepackage{tabularx}
\usepackage{subfigure}
\usepackage{color}
\usepackage{cite}
\usepackage{enumitem}

\usepackage[T1]{fontenc}
\usepackage{amsmath}
\usepackage[cmintegrals]{newtxmath}
\usepackage{bm}

\usepackage{multirow}

\hypersetup{hidelinks}

\newtheorem{definition}{Definition}
\newtheorem{remark}{Remark}

\definecolor{airforceblue}{rgb}{0.36, 0.54, 0.66}
\definecolor{applegreen}{rgb}{0.55, 0.71, 0.0}
\definecolor{bittersweet}{rgb}{1.0, 0.44, 0.37}

\LetLtxMacro{\originaleqref}{\eqref}

\renewcommand{\eqref}{\originaleqref}

\newcommand{\rv}{\color{black}}

\begin{document}

\title{Distribution Locational Marginal Emission for Carbon Alleviation in Distribution Networks: Formulation, Calculation, and Implication}

\author{Linwei~Sang, 
Yinliang~Xu,~\IEEEmembership{Senior Member,~IEEE}, 
Hongbin~Sun,~\IEEEmembership{Fellow,~IEEE}, \\
Qiuwei Wu,~\IEEEmembership{Senior Member,~IEEE},
and Wenchuan~Wu,~\IEEEmembership{Fellow,~IEEE}.
\thanks{Linwei~Sang, Yinliang~Xu, and Qiuwei Wu are with Tsinghua-Berkeley Shenzhen Institute, Tsinghua Shenzhen International Graduate School, Tsinghua University, Shenzhen, China. (E-mail: \url{sanglinwei@berkeley.edu}, \url{xu.yinliang@sz.tsinghua.edu.cn}, \url{qiuwu@sz.tsinghua.edu.cn}.)}
\thanks{Wenchuan~Wu and Hongbin~Sun are with the Department of Electrical Engineering, Tsinghua University, Beijing, China.}
}

\maketitle

\begin{abstract}

Regulating the proper carbon-aware intervention policy is one of the keys to emission alleviation in the distribution network, whose basis lies in effectively attributing the emission responsibility using emission factors. This paper establishes the distribution locational marginal emission (DLME) to calculate the marginal change of emission from the marginal change of both active and reactive load demand for incentivizing carbon alleviation. It first formulates the day-head distribution network scheduling model based on the second-order cone program (SOCP). The emission propagation and responsibility are analyzed from demand to supply to system emission. Considering the complex and implicit mapping of the SOCP-based scheduling model, the implicit theorem is leveraged to exploit the optimal condition of SOCP. The corresponding SOCP-based implicit derivation approach is proposed to calculate the DLMEs effectively in a model-based way. Comprehensive numerical studies are conducted to verify the superiority of the proposed method by comparing its calculation efficacy to the conventional marginal estimation approach, assessing its effectiveness in carbon alleviation with comparison to the average emission factors, and evaluating its carbon alleviation ability of reactive DLME.

\end{abstract}

\begin{IEEEkeywords}
Emission factor, second-order cone program, distributed energy resources, implicit mapping, distribution locational marginal emission.
\end{IEEEkeywords}

\section{Introduction}

\IEEEPARstart{D}{ecarbonizing} the electrical energy system is urgent and essential to tackle the increasingly visible climate changes \cite{CHEN20212715}. Accelerating the carbon-neutral transition requires massive renewables penetrated in the distribution networks and the corresponding facilitation from technology, market, and policy areas \cite{XIE2022100640}. Regulating the proper carbon-aware intervention policy is one of the keys to emission alleviation \cite{Donti2019, Wang2021}, and its basis lies in attributing the emission responsibility from the power supply to the load demand in the electrical power networks temporally and spatially using emission factors. 

Current emission factors can be categorized into average emission factors (AEFs) and marginal emission factors (MEFs) \cite{Donti2019, Valenzuela2023}. AEFs quantify the average emission intensity per MWh of load consumption and can be calculated by the popular carbon emission flow (CEF) models \cite{Kang2015, Sang2023}, which measure emissions over long periods; in contrast, MEFs quantify the ratio of the change of emission intensity to the marginal change of load demand \cite{WANG2014141, PARK2023120303}, which measures emission impacts to small local load change. Among the MEFs, locational marginal emissions (LMEs) are proposed in \cite{Donti2019, Valenzuela2023, Deetjen2019} to quantify the emission sensitivity at the nodal level, indicating the spatial heterogeneity in marginal emission rates that emerge from network constraints for transmission networks. LMEs can be calculated by the regression-based and model-based approaches. Regression-based approaches utilize the emission and load data to estimate the values of LME across the regions \cite{Donti2019, HAWKES20105977}, which require plenty of existing data. In contrast, model-based approaches utilize the transmission dispatching model to identify the marginal generators, whose emissions are the marginal emissions \cite{Deetjen2019, WANG2014141}. 

Intuitively, LMEs are the emission equivalent of locational marginal prices (LMPs) in the power systems \cite{Valenzuela2023}. LMPs have been studied and applied extensively to incentivize the economic operation of the transmission system \cite{Antonio2023}, and, similarly, LMEs can incentivize the low-carbon operation of the transmission network. With the increase of renewables and demand flexibility in distribution networks, distribution locational marginal price (DLMP) is proposed in \cite{Liu2018, Bai2018, Deng2019, Zhao2023} to extend the LMPs for distribution network management. Liu \emph{et al.} \cite{Liu2018} propose the integrated DLMP for optimal electrical vehicle management based on DC optimal power flow (OPF). Bai \emph{et al.} \cite{Bai2018} leverage the linearized AC OPF to derive the DLMPs for the active/reactive power and facilitate congestion management and voltage support in day-ahead distribution market clearing. Unlike the transmission network, the distribution network with the high R/X ratio cannot neglect the power loss in system operation, which requires more precise power flow models, e.g., the second-order cone program (SOCP) formulation \cite{Sang2022}. These characteristics lead to additional emissions from reactive load demand in the distribution networks. The main reason is that, though generating reactive power does not generate emissions directly, delivering reactive power will incur active power loss, leading to additional emissions beyond the emissions from active load demand. The above distribution network characteristics raise the challenges of quantifying the emission factors for load demand temporally and spatially to alleviate carbon emission effectively. 

In analogy to extending LMPs to DLMPs, this paper extends the LME to establish the distribution location marginal emissions (DLMEs) for active and reactive power to incentivize carbon alleviation in the distribution network. It formulates the SOCP-based distribution network scheduling model, analyzes the emission propagation and responsibility from demand to generation to system emission, and leverages the SOCP-based implicit derivation approach to calculate the DLMEs effectively in a model-based way. Comprehensive numerical studies are conducted to validate the proposed method, including comparing its calculation efficacy to the conventional LME estimation approach, assessing its effectiveness in carbon alleviation with comparison to the AEFs from CEF, and evaluating its carbon alleviation ability of reactive DLME. So the main contributions of this paper can be summarized as follows:

1) \emph{Formulation:} Inspired by DLMP, we, \emph{firstly}, formulate the distribution locational marginal emission (DLME) for both active and reactive power to incentivize the carbon alleviation in the SOCP-based distribution scheduling model.

2) \emph{Calculation:} Based on optimal conditions of SOCP, we leverage the implicit derivation of the SOCP solution mapping to calculate the DLME of distribution networks in a model-based manner via emission propagation analysis and emission responsibility attribution.

3) \emph{Implication:} Case studies are conducted to validate the advantages of DLME by \emph{i)} its calculation efficacy over the conventional LME estimation approach, \emph{ii)} its carbon alleviation effectiveness over comparison to the AEFs from CEF, \emph{iii)} and its carbon alleviation ability of reactive DLME.

Significant differences can be observed when comparing this work with previous work \cite{Sang2022}. \textit{i)} they focus on different problems: the former regulates the carbon-aware intervention policy for the distribution system, and the latter regulates the voltage devices; \textit{ii)} they have different goals: the former aims to incentivize the carbon alleviation effectively, and the latter achieves the system safe and economic operation;  \textit{iii)} they are solved by different techniques: the former is solved in the one step backward way, and the latter is solved in a stochastic gradient descent way.

The rest of this paper is organized as follows. Section \ref{sec: prob} prepares the SOCP-based distribution network scheduling model with emerging distributed energy resources formulation and develops the distribution locational marginal emission formulation. Section \ref{sec: method} proposes the differential DLME calculation approach via the emission propagation analysis and implicit gradient derivation. Section \ref{sec: case} verifies the effectiveness of the proposed formulation and calculation approach. Section \ref{sec: conclude} concludes this paper.

\section{Distribution Locational Marginal Emission Formulation} \label{sec: prob}

This section firstly prepares the emerging distributed energy resources formulation in \ref{sub: der} and the corresponding SOCP-based distribution network scheduling model in \ref{sub: scheduling}. Then, based on the above, it proposes the DLME formulation by the emission propagation and responsibility analysis in \ref{sub: DLME}.

\subsection{Distributed Energy Resources Formulation} \label{sub: der}

We consider comprehensive distributed energy resource (DER) formulation in the SOCP-based distribution network scheduling model, including the distributed generators, distributed energy resources, and the emerging EV aggregator model sequentially.

\subsubsection{Distributed Generators}

We formulate the distributed generators (DGs) with two common forms the emerging inverter-based generators of \eqref{eq: ib dg} and the conventional synchronous machine-based generators of \eqref{eq: sm dg}.

\paragraph{Inverter-based DG} 

Distributed wind turbines (WTs) and photovoltaics (PVs) are connected to the distribution network through the power electronics inverters to provide the active power supply and reactive power support. Due to technical requirements, the power factors of the inverters are limited to a given range in \eqref{eq: ib dg coef}.

\begin{subequations}
  \allowdisplaybreaks
  \begin{align}
    &P^{inv}_{i,t,\min} \leq p^{inv}_{i,t} \leq P^{inv}_{i,t,\max}, &&\forall i\in\mathcal{IB}, \forall t\in\mathcal{T} \label{eq: ib dg limit}  \\ 
    &\frac{p^{inv}_{i,t}}{\sqrt{(p^{inv}_{i,t})^2 + (q^{inv}_{i,t})^2}}\geq \kappa_{i, \min}, &&\forall i\in\mathcal{IB}, \forall t\in\mathcal{T} \label{eq: ib dg coef} 
  \end{align}
  \label{eq: ib dg}
\end{subequations}
where $p^{inv}_{i,t}$ and $q^{inv}_{i,t}$ are the active and reactive power of inverter-based DG $i$ in time $t$; $\kappa_{i, \min}$ is the minimum power factor of inverter-based DG $i$; $P^{inv}_{i,t,\min}$ and $P^{inv}_{i,t,\min}$ are the minimum and maximum active power of inverter-based DG $i$ in time $t$.

The complex \eqref{eq: ib dg coef} can be further transformed into the linear constraints in \eqref{eq: ib dg trans}.

\begin{eqnarray}
  \begin{aligned}
    & -\frac{p^{inv}_{i,t}\sqrt{1-(\kappa_{i, \min})^2}}{\kappa_{i, \min}} \leq q^{inv}_{i,t} \leq \frac{p^{inv}_{i,t}\sqrt{1-(\kappa_{i, \min})^2}}{\kappa_{i, \min}}.
  \end{aligned}
  \label{eq: ib dg trans}
\end{eqnarray}

The maximum outputs of the above inverter-based DGs mainly depend on the solar radiation for PVs and the wind speed for WTs in \eqref{eq: ib dg limit}.

\paragraph{Synchronous machine-based DG}

Distributed synchronous generators rely on conventional fossil-based energy resources for the active power supply and the excitation for the reactive power supply for the distribution network in \eqref{eq: sm dg}. Unlike the inverter-based DG, synchronous generators' active power generation is further limited by their physical constraints, as formulated as the ramping constraints in \eqref{eq: sm dg ramp}.

\begin{subequations}
  \begin{align}
    & P^{sm}_{i,t,\min} \leq p^{sm}_{i,t} \leq P^{sm}_{i,t,\max}, &&\forall i\in\mathcal{SM}, \forall t\in\mathcal{T} \label{eq: sm dg p} \\ 
    & Q^{sm}_{i,t,\min} \leq q^{sm}_{i,t} \leq Q^{sm}_{i,t,\max}, &&\forall i\in\mathcal{SM}, \forall t\in\mathcal{T} \label{eq: sm dg q} \\ 
    & r^{RD}_{i} \leq P^{sm}_{i,t} - P^{sm}_{i,t-1} \leq r^{RU}_{i} &&\forall i\in\mathcal{SM}, \forall t\in\mathcal{T} \label{eq: sm dg ramp}
  \end{align}
  \label{eq: sm dg}
\end{subequations}
where $p^{sm}_{i,t}$ and $q^{sm}_{i,t}$ are the active and reactive power of synchronous machine-based DG $i$ in time $t$; $P^{sm}_{i,t,\min}$ and $P^{sm}_{i,t,\max}$ are the minimum and maximum active power; $Q^{sm}_{i,t,\min}$ and $Q^{sm}_{i,t,\max}$ are the minimum and maximum reactive power; $r^{RD}_{i}$ and $r^{RU}_{i}$ are the maximum ramping down and ramping up values of synchronous machine $i$; $\mathcal{SM}$ is the set of synchronous machine-based DGs.

\subsubsection{Distributed Energy Storage Model}

Distributed energy storage (ES) models mainly provide the active power for temporal energy transferring flexibility with the storage capacity, power, and energy change limits in \eqref{eq: es}.

\begin{subequations}
  \allowdisplaybreaks
  \small
  \begin{align}
    & 0 \leq p^{cha}_{i,t} \leq P^{cha}_{i,\max}, 0 \leq p^{dis}_{i,t} \leq P^{dis}_{i, \max}, && \forall i\in\mathcal{ES}, \forall t\in\mathcal{T} \label{eq: es power}\\
    & e_{i,t} = e_{i,t-1} + \eta^{cha}_{i} p^{cha}_{i,t} - (1/\eta^{dis}_{i}) p^{dis}_{i,t}, && \forall i\in\mathcal{ES}, \forall t\in\mathcal{T}  \label{eq: es state} \\
    & E_{i, \min} \leq e_{i,t} \leq E_{i, \max}, && \forall i\in\mathcal{ES}, \forall t\in\mathcal{T} \label{eq: es capacity} 
  \end{align}
  \label{eq: es}
\end{subequations}
where $p^{cha}_{i,t}$ and $p^{dis}_{i,t}$ are the charging and discharging power of ES $i$ in time $t$; $e_{i, t}$ is the stored energy; $P^{cha}_{i,\max}$ and $P^{dis}_{i, \max}$ are the maximum charging and discharging power; $E_{i, \min}$ and $E_{i, \max}$ are the minimum and maximum stored energy; $\eta^{cha}_{i}$ and $\eta^{dis}_{i}$ are the charging and discharging coefficients.

To minimize operation cost, the simultaneous charging and discharging of ES in \eqref{eq: es power} will not coincide at the same time due to the power losses in the charging and discharging, so binary variables indicating charging/discharging status can be eliminated in \eqref{eq: es state}.

\subsubsection{EV Aggregators with Flexibility}

Based on \cite{Zhang2017}, the aggregated charging power of electrical vehicle (EV) aggregators can provide temporal load demand flexibility under the satisfaction of each EV's energy demand with the power bounds constraint of \eqref{eq: ev power} and the energy bound constraint of \eqref{eq: ev energy}.

\begin{subequations}
  \allowdisplaybreaks
  \begin{align}
    & p^{EV}_{i, t, lb} \leq p^{EV}_{i, t} \leq p^{EV}_{i, t, ub}, && \forall i\in\mathcal{EV}, \forall t\in\mathcal{T} \label{eq: ev power} \\ 
    & e^{EV}_{i, t, lb} \leq \sum^{t}_{\tau=1} p^{EV}_{\tau, t} \leq e^{EV}_{i, t, ub}, && \forall i\in\mathcal{EV}, \forall t\in\mathcal{T} \label{eq: ev energy}
  \end{align}
  \label{eq: ev}
\end{subequations}
where $p^{EV}_{i, t}$ is the active power of EV fleet $i$ in time $t$; $p^{EV}_{i, t, lb}$ and $p^{EV}_{i, t, ub}$ are the corresponding low bound and upper power bound; $e^{EV}_{i, t, lb}$ and $e^{EV}_{i, t, ub}$ are the corresponding low bound and upper energy bound. 

\subsection{SOCP-based Scheduling Formulation} \label{sub: scheduling}

On top of the DERs formulation, we formulate a SOCP-based day-ahead distribution network management model in \eqref{eq: dso da}. The main difference between the real-time and day-ahead models lies in considering the temporal constraints of synchronous machine-based distributed generators, distributed energy storage, and EV aggregators. 

\subsubsection{Day-ahead Scheduling Model}

SOCP-based day-ahead scheduling model is formulated in \eqref{eq: dso da}, which considers the temporal relationship of energy devices with fossil-based generator ramping and energy storage coupling constraints. The corresponding DLME from the day-ahead scheduling model reflects the temporal and spatial emission factors.

\begin{subequations}
  \small
  \allowdisplaybreaks
  \begin{align}
    \min_{q^{reg}} & \sum_{t\in \mathcal{T}} \Bigl\{\sigma^{P}_{S,t}p^{S}_{t} + \sigma^{Q}_{S,t}q^{S}_{t} + \sum_{i\in\mathcal{DER}}(\sigma^{P}_{S,i}p^{G}_{i, t} + \sigma^{Q}_{S,t}q^{G}_{i, t})  \Bigr\} \label{eq: da obj}  \\
    & \mathcal{DER} = \{\mathcal{IB}, \mathcal{SM}, \mathcal{ES} \} \notag \\  
    \text{s.t.} \quad &\sum_{jk \in \mathcal{E}} p_{jk,t} - \sum_{ij \in \mathcal{E}} (p_{ij, t} - r_{ij}l_{ij,t}) = {p}^{G}_{j,t} - {p}^{D}_{j,t} \label{eq: da pf p}\\
    & \sum_{jk \in \mathcal{E}} q_{jk,t} - \sum_{ij \in \mathcal{E}} (q_{ij, t} - x_{ij}l_{ij,t}) = {q}^{G}_{j,t} - {q}^{D}_{j,t} \label{eq: socp c}\\
    & v_{j,t} = v_{i,t} + (r^2_{ij}+x^2_{ij})l_{ij,t} - 2 (r_{ij}p_{ij,t} + x_{ij}q_{ij,t}) \forall ij \in \mathcal{E} \label{eq: socp d}\\
    & || 2p_{ij,t} \quad 2q_{ij,t} \quad l_{ij,t}-v_{i,t} || \leq l_{ij,t} + v_{i,t} \label{eq: socp e}\\
    & (V^{\min}_{i,t})^2 \leq v_{i,t} \leq (V^{\max}_{i,t})^2, \forall i \in \mathcal{N}_B \label{eq: socp f}\\
    & (I^{\min}_{ij,t})^2 \leq l_{ij,t} \leq (I^{\max}_{ij,t})^2, \forall ij \in \mathcal{E} \label{eq: socp g} \\ 
    & \text{DER constraints:} \eqref{eq: ib dg}, \eqref{eq: sm dg}, \eqref{eq: es}, \eqref{eq: ev}
  \end{align}
  \label{eq: dso da}
\end{subequations}
where ${p}^{S}_{j,t}$ and ${q}^{S}_{j,t}$ are the active and reactive power supply at substation; ${p}^{G}_{j,t}$ and ${q}^{G}_{j,t}$ are the active and reactive power supply of DERs; ${p}^{D}_{j,t}$ and ${q}^{D}_{j,t}$ are the active and reactive bus load demand; $v_{i,t}$ and $l_{ij,t}$ are the square of voltage magnitude and current; $\sigma^{P}_{S,t}$ and $\sigma^{Q}_{S,t}$ are the active and reactive prices of the substation; $\sigma^{P}_{G,t}$ and $\sigma^{Q}_{G,t}$ are the active and reactive prices of DERs; $r_{ij}$ and $x_{ij}$ are the resistance and reactance of branch $ij$;  $p_{ij,t}$ and $q_{ij,t}$ are the active and reactive power of branches; $\mathcal{T}$ and $\mathcal{E}$ are the set of operation periods and network branches. Equations \eqref{eq: da pf p} and \eqref{eq: socp c} formulate the bus active and reactive balance, \eqref{eq: socp e} and  \eqref{eq: socp e} formulate the branch model and relax the nonlinear power flow equations by the second-order cone constraints, and \eqref{eq: socp f} and \eqref{eq: socp g} limit the voltage and branch power transmission. The detailed formulation is shown in \cite{Bai2018}.

\subsubsection{Scheduling Solution Mapping}

The above scheduling model \eqref{eq: dso da} maps from the load demand $D = (p_{d}, q_{d})\in\mathbb{R}^{2N_{B}\times T}$ to the optimal power supply $G = (p_{g}, q_{g})\in\mathbb{R}^{2N_{G}\times T}$ implicitly under the optimal solution of SOCP. Let $(p_{g}, q_{g}) = f_{g} (D):\mathbb{R}^{2N_{B}\times T}\rightarrow \mathbb{R}^{2N_{G}\times T}$ denote the above mapping, called the \emph{scheduling solution mapping}. 

\begin{remark}
  For discrete variables of SOCP-based scheduling, such as the feeder status, we first run the mixed-integer second order-cone programming (MISOCP)-based model to determine the value of integer variables and then fix the integer value to formulate the SOCP-based solution mapping for DLME calculation in the proposed differential approach.
\end{remark}

\subsection{Distribution Locational Marginal Emission Formulation} \label{sub: DLME}

The day-ahead SOCP-based scheduling model \eqref{eq: dso da} achieves the optimal active/reactive power allocation under the security requirement in the distribution networks. At the same time, the emissions are generated, including the bulk and the DER generation emission. Let $e_{g}$ denote the emission of generator $g$, which is a function of $p_g$, denoted by $f_{e}(p_{g})$. Then we analyze the emission from the demand side to the generation side in section~\ref{subsub: forward}, allocate the emission responsibility from the generation side to the demand side in section~\ref{subsub: back}, and marginally formulate the DLME in section~\ref{subsub: forward}.

\subsubsection{Emission Propagation Analysis} \label{subsub: forward}

For each time $t$, in the emission propagation of \eqref{eq: forward}, the active/reactive power $(p_{d}, q_{d})$ in the load demand requires the distributed generator energy supply $(p_{g}, q_{g})$ under the operational requirements via the optimal distribution network management \eqref{eq: dso da},  denoted by \emph{scheduling solution mapping}, in \eqref{eq: fg}, $(p_{g}, q_{g})$ leads to the generator emission $e_g$ in \eqref{eq: fe}, and the total emission $e_{sum}$ can be obtained by accumulating each generator's emission in \eqref{eq: esum}.

\begin{subequations}
  \allowdisplaybreaks
  \begin{align}
    & {(p_{d}, q_{d}) \Rightarrow (p_{g}, q_{g})}  \Rightarrow e_{g} \Rightarrow e_{sum} \\ 
    & \text{Scheduling solution mapping: } (p_{g}, q_{g}) = f_{g} (p_{d}, q_{d}) \label{eq: fg} \\
    & \text{Emission mapping: } e_{g} = f_{e}(p_{g}), g\in\mathcal{G} \label{eq: fe} \\ 
    & \text{Emission summing: }e_{sum} = \sum_{g\in\mathcal{G}} e_{g} \label{eq: esum}
  \end{align}
  \label{eq: forward}
\end{subequations}

\subsubsection{Emission Responsibility Analysis}\label{subsub: back}

Load demand should take responsibility for the distribution network emission by the corresponding emission factors. We consider the emission factors in the distribution network from the marginal perspective. The slight change in active/reactive load demand will lead to change the distribution system emission change, denoted by the marginal emission. Current LME calculation focuses on the transmission network ignoring the power loss, which leverages the DC optimal power flow model for the model-based LME calculation in \cite{Alberto2023}. We borrow the idea of extending LMP to DLMP for the LME in the distribution network. Unlike the DC power flow-based LME, for the distribution network of \eqref{eq: dso da}, the active and reactive are coupled by the SOCP-based power flow model, and reactive power will induce active power flow loss to generate system emission. Active power generation of fossil-based will generate the system emissions.

\subsubsection{Distribution Locational Marginal Emission} \label{subsub: dlme}

So, based on the above, we quantify the load emission factors for carbon alleviation in the distribution networks by DLME as follows:
\begin{definition}[Distribution locational marginal emission]
  Distribution locational marginal emission (DLME) is formulated as the ratio of the marginal change of the total emission in the distribution network emission with respect to marginal active and reactive load demand change, as presented in \eqref{eq: DLME def} and \eqref{eq: DLME reactive}.
\end{definition}
\begin{eqnarray}
  \begin{aligned}
    \text{DLME} = \frac{\partial e_{sum}}{\partial p_{d}} = \lim_{\Delta_p\rightarrow 0} \frac{e_{sum}(p_{d}, q_{d}) - e_{sum}(p_{d}- \Delta_{p}, q_{d})}{\Delta_{p}}
  \end{aligned}
  \label{eq: DLME def}
\end{eqnarray}
\begin{eqnarray}
  \begin{aligned}
    \text{DLME}_{q} = \frac{\partial e_{sum}}{\partial q_{d}} = \lim_{\Delta_q\rightarrow 0} \frac{e_{sum}(p_{d}, q_{d}) - e_{sum}(p_{d}, q_{d}- \Delta_{q})}{\Delta_{q}}
  \end{aligned}
  \label{eq: DLME reactive}
\end{eqnarray}
We note that $\text{DLME}$ without subscript denotes active DLME for active load demand, and $\text{DLME}_{q}$ denotes the reactive DLME for reactive load demand. The relationship between the $e_{sum}$ and $p_{d}$/$q_{d}$ is analyzed with the mapping of $f_{g}(\cdot)$ and $f_{e}(\cdot)$ in \eqref{eq: forward}. $\text{DLME}$ and $\text{DLME}_{q}$ cannot be calculated directly and explicitly due to the complex composite relationship.

\section{Methodology} \label{sec: method}

This section presents the calculation of DLME through the implicit derivation approach by introducing the overall calculation framework in section \ref{sub: frame}, analyzing the backward emission responsibility in section \ref{sub: back}, and providing the solution mapping with the derivative of SOCP in section \ref{sub: socp map}. We note that the following calculation analysis focuses on the $\text{DLME}$, and $\text{DLME}_{q}$ can be calculated in the same way.

\subsection{Overall Framework of the DLME Calculation} \label{sub: frame}

\begin{figure}[ht]
  \centering
  \rv
  \includegraphics[scale=1.3]{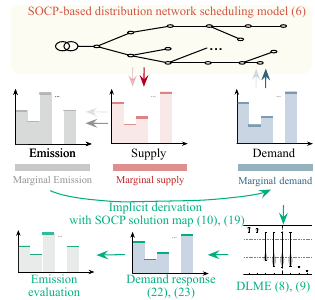}
  \caption{Overall framework of the DLME framework.}
  \label{fig: DLME framework}
\end{figure}

Based on the DLME formulation of section~\ref{sub: DLME}, we formulate the overall framework of the DLME calculation in the emission propagation and responsibility analysis in Fig.~\ref{fig: DLME framework}. 
Marginal demand leads to marginal power supply, and marginal supply further leads to the system's marginal emission (from right to left); DLME is the composite of the derivatives of the emission to the supply and the derivatives of the supply to the demand (from left to right) in Fig.~\ref{fig: DLME framework}. The calculated DLME can incentivize effective demand response for carbon alleviation in the bottom of Fig.~\ref{fig: DLME framework}.

\subsection{Backward Emission Responsibility Derivation} \label{sub: back}

The key of DLME lies in computing the derivative of $e_{sum}$ concerning $p_{d}$ in \eqref{eq: backward}. Based on the emission propagation process of \eqref{eq: forward} and the chain rules, we decompose the composite derivative calculation into three parts differentiating through total emission \eqref{eq: grad e_sum}, through each node emission \eqref{eq: grad e_g}, and power generation \eqref{eq: grad p_g}.

\begin{subequations}
  \allowdisplaybreaks
  \begin{align}
    &\frac{\partial e_{sum}}{\partial p_{d}} = 
    \frac{\partial e_{sum}}{\partial e_{g}}
    \frac{\partial e_{g}}{\partial p_{g}} \frac{\partial p_{g}}{\partial p_{d}} \label{eq: grad composite} \\ 
    &\text{Differentiate through total emission:} \frac{\partial e_{sum}}{\partial e_{g}} \label{eq: grad e_sum} \\
    &\text{Differentiate through each emission:} \frac{\partial e_{g}}{\partial (p_{g}, q_{g})} \label{eq: grad e_g}\\
    &\text{Differentiate through power generation:} \frac{\partial p_{g}}{\partial (p_{d}, q_{d})} \label{eq: grad p_g}
  \end{align}
  \label{eq: backward}
\end{subequations}

The differentiation calculation of \eqref{eq: grad e_sum} and \eqref{eq: grad e_g} relies on \eqref{eq: esum} and \eqref{eq: fe}, which features the explicit expressions; in contrast, the calculation of \eqref{eq: grad p_g} relies on the optimal SOCP-based scheduling of \eqref{eq: fg}, which the derivatives are embedded in the optimal condition in an implicit form, leading to the main challenge of the differential DLME calculation approach.

\subsection{The Solution Mapping with Its Derivative of SOCP via Implicit Derivation}\label{sub: socp map}

\begin{figure}[ht]
  \centering
  \includegraphics[scale=0.85]{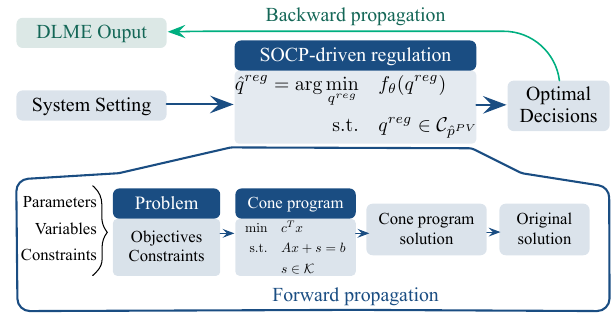}
  \caption{The solution map and its derivative of the general SOCP for DLME calculation.}
  \label{fig: cvxlayers}
\end{figure}

The main challenge of \eqref{eq: backward} lies in the differentiating through the distribution network scheduling model of \eqref{eq: grad p_g}. We propose to leverage the \emph{implicit function theorem} from \cite{dontchev2014implicit} to calculate the derivative of \eqref{eq: grad p_g} through the SOCP differentiation. Calculating \eqref{eq: grad p_g} cannot be implemented explicitly due to the complex mapping relationship of \eqref{eq: dso da}. But it is possible to differentiate through the SOCP-based scheduling model by implicitly differentiating its optimality conditions. The main scheme of the SOCP solution mapping and its derivative comprises the forward propagation for optimal decision and the backward derivatives for LME, as presented in Fig.~\ref{fig: cvxlayers}.

The parameterized SOCP-based \emph{scheduling solution mapping} is decomposed into three stages: an affine mapping from input parameters to the skew matrix, a solver, and an affine mapping from the solver solution to the original problem solution. Then the gradients from the solution to the input parameters can be back-propagated based on the chain rule. The rest elaborates on the above three components under a general convex conic program formulation.

Consider the general form of the SOCP-based scheduling model of \eqref{eq: dso da} as a conic program with the primal and dual forms:
\begin{equation}
  \begin{aligned}[c]
    (\text{Primal}) \min \text{ }  & c^T x \\
    \text{s.t.} \text{ } & A x + s = b \\
    & s \in \mathcal{K} \\
  \end{aligned}
  \quad
  \begin{aligned}[c]
    (\text{Dual}) \min \text{ } & b^T y \\
    \text{s.t.} \text{ } & A^T y + c = 0 \\
    & y \in \mathcal{K}^* 
  \end{aligned}
  \label{eq: general socp}
\end{equation}
where $x\in\mathbb{R}^n$ and $y\in\mathbb{R}^m$ are the primal and dual variables of \eqref{eq: dso da}, where $p_{g}$ is included in $x$ and $(p_{d}, q_{d})$ is included in $b$; $s$ is the primal slack variable; $\mathcal{K}\in\mathbb{R}^m$ is the closed, convex cone with its dual cone $\mathcal{K}^*\in\mathbb{R}^m$, including the second-order cone constraints of the distribution power flow and distribution system operational constraints.

According to the Karush-Kuhn-Tucker (KKT) conditions, the optimal $(x^*, y^*, s^*)$ of \eqref{eq: general socp} should satisfy:
\begin{subequations}
  \allowdisplaybreaks
  \begin{align}
    Ax + s = b, A^T y + c = 0 \\
    s \in \mathcal{K}, y \in \mathcal{K}^*, s^T y = 0 .
  \end{align}
  \label{eq: optimal condition}
\end{subequations}

The solution mapping from the input parameters $(A, b, c)$ to the distribution network optimal solution $(x^*, y^*, s^*)$ is formulated as $\mathcal{F}: \mathbb{R}^{m\times n} \times \mathbb{R}^m \times \mathbb{R}^n \to \mathbb{R}^{n+2m}$, and then the $\mathcal{F}$ is decomposed  into $\phi \circ s \circ Q$, where $\circ$ is the function composition operator\footnote{For example, $h = f \circ g$ refers to $h(x) = f(g(x))$}. Specifically, $Q$ maps the input parameters $(A, b, c)$ to the corresponding skew-symmetric matrix $Q$; $s$ takes $Q$ as input and solves the homogeneous self-dual embedding for implicit differentiation with the intermediate output variable $z$; $\phi$ maps the $z$ to the optimal primal-dual solution $(x^*, y^*, s^*)$.

Based on the composition, the corresponding total derivative of the solution mapping $\mathcal{F}$ is derived as: 
\begin{equation}
  \mathsf{D} \mathcal{F} (A, b, c) = \mathsf{D} \phi (z) \mathsf{D} s(Q) \mathsf{D} Q(A, b, c)
  \label{eq: solution map}
\end{equation}
where $\mathsf{D}$ is the derivative operator, and $\mathsf{D}f(x)$ is the derivative of function $f$ with respect to $x$.

\paragraph{Skew-symmetric Mapping}\label{para: D a}
The skew-symmetric mapping defines the Q matrix as:
\begin{equation}
  Q(A, b ,c) =
  \begin{bmatrix}
    0 & A^T & c \\
    -A & 0 & b \\
    -c^T & -b^T & 0 \\
  \end{bmatrix} \in \mathcal{Q}.
  \label{eq: Q matrix}
\end{equation}
The corresponding derivative of $Q(A, b, c)$ with respect to the $(A, b, c)$ is formulated as:
\begin{equation}
  \mathsf{D} Q(A, b ,c) =
  \begin{bmatrix}
    0 & \mathsf{d} A^T & \mathsf{d} c \\
    -\mathsf{d} A & 0 & \mathsf{d} b \\
    -\mathsf{d} c^T & -\mathsf{d}  b^T & 0 \\
  \end{bmatrix}.
  \label{eq: Q derivative}
\end{equation}

\paragraph{Homogeneous Self-dual Embedding}\label{para: D b}

The homogeneous self-dual embedding finds a zero point of a certain residual mapping by solving \eqref{eq: general socp}. It utilizes the variable $z$ as an intermediate variable, partitioned as $(u, v, \omega)$, to formulate the normalized residual map function from \cite{Busseti2019} as:
\begin{equation}
  \mathcal{N}(z, Q) = ((Q - I)\Pi + I) (z / |\omega|)
  \label{eq: homogeneous}
\end{equation}
where $\Pi$ is the project operation onto $\mathbb{R} \times \mathcal{K} \times R_{+}$.

If and only if $\mathcal{N}(z, Q) = 0$ and $\omega >0$, the $z$ can construct the solution of \eqref{eq: general socp} for given $Q$. Then we derive the derivatives of $\mathcal{N}(z, Q)$ with respect to $z$ and $Q$:
\begin{subequations}
  \begin{align}
    \mathsf{D}_{Q} \mathcal{N}(z, Q) =& \Pi(z/|\omega|) \\
    \mathsf{D}_{z} \mathcal{N}(z, Q) 
    =& ((Q-I)\mathsf{D}\Pi(z)+I)/|\omega| \nonumber \\
    &- \text{sign}(\omega)((Q-I)\Pi + I)(z/\omega^2) e^T  \\
    =& ((Q-I)\mathsf{D}\Pi(z)+I)/\omega \\
    &\text{ when $z$ is the solution of \eqref{eq: general socp} and $\omega > 0$} \nonumber
  \end{align}
  \label{eq: DzN}
\end{subequations}
where $e \in \mathbb{R}^n$ is $(0,0, ..., 1)$.

\begin{definition}[Implicit function theorem, \cite{dontchev2014implicit}]
  For a given $Q$, $z$ is the solution of the primal-dual pair problem of \eqref{eq: general socp}, and $\Pi$ is differentiable at $z$. Then $\mathcal{N}$ is differentiable at $z$ with $\mathcal{N}(z, Q)=0$ and $\omega > 0$.  
\end{definition}

Based on the above implicit function theorem, there exists a neighborhood $V \subseteq \mathcal{Q}$ such that the $z=s(Q)$ of $\mathcal{N}(z, Q)$ is unique. So $\mathsf{D}_Q\mathcal{N}(z(Q), Q)$ is zero when $z$ is the solution of \eqref{eq: general socp}. Then the derivative of $s(Q)$ with respect to Q is derived based on the derivatives with respect to $z$ and $Q$ \eqref{eq: DzN} as:

\begin{subequations}
  \allowdisplaybreaks
  \begin{align}
    & \mathsf{D} \mathcal{N}_{z}(s(Q, Q) \mathsf{D} s(Q) +  \mathsf{D}_{Q} \mathcal{N}(z, Q) = 0 \\
    \Rightarrow & \mathsf{D} s(Q) = - \mathsf{D}_{z} \mathcal{N}_{z}(s(Q, Q)^{-1} \mathsf{D}_{Q} \mathcal{N}(z, Q)).
  \end{align}
  \label{eq: DsQ}
\end{subequations}

\paragraph{Solution Construction}\label{para: D c}

The solution construction function $\phi(z)$ construct the optimal solution $(x^*, y^*, s^*)\in\mathbb{R}^{n+2m}$ from the primal-dual pair \eqref{eq: general socp} from $z=(u, v, \omega) \in \mathbb{R}^N$ as:
\begin{equation}
  (x^*, y^*, s^*) = \phi(z) = (u, \Pi_{\mathcal{K}^*}(v), \Pi_{\mathcal{K}^*}(v) -v) / \omega
  \label{eq: solution constr}
\end{equation}
where $\Pi_{\mathcal{K}^*}$ is projection operator onto the dual cone $\mathcal{K}^*$.

So the corresponding $\phi(z)$ is differentiable with the derivatives as

\begin{equation}
  \mathsf{D} \phi (z) =
  \begin{bmatrix}
    I & 0 & -x^{*} \\
    0 & \mathsf{D} \Pi_{\mathcal{K}^*}(v) & -y^{*} \\
    0 & \mathsf{D} \Pi_{\mathcal{K}^*}(v) - I^{*} & -s^{*} \\ 
  \end{bmatrix}/\omega.
  \label{eq: phi z}
\end{equation}

Applying the derivative of $\mathsf{D}\mathcal{F}(A, b, c)$ to the perturbation of $(\mathsf{d}A, \mathsf{d}b, \mathsf{d}c)$ is formulated by stacking the above three derivative components of \eqref{eq: Q derivative}, \eqref{eq: DsQ}, and \eqref{eq: phi z}:
\begin{eqnarray}
  \begin{aligned}
    (\mathsf{d}x^{*}, \mathsf{d}y^{*}, \mathsf{d}s^{*}) = \mathsf{D}\mathcal{F}(A, b, c) (\mathsf{d}A, \mathsf{d}b, \mathsf{d}c)
  \end{aligned}
  \label{eq: apply}
\end{eqnarray}
The above implicit derivative of the SOCP solution calculation method \eqref{eq: apply} can obtain the backward propagation derivatives from the implicit function theorem for the derivation of \eqref{eq: grad p_g}.

\section{Case Study} \label{sec: case}

We utilize the modified IEEE 33-bus distribution networks with multiple PVs to verify the carbon alleviation effectiveness of the proposed DLME and the calculation efficacy of the corresponding implicit derivation calculation approach. We construct the day-ahead SOCP-based scheduling models of \eqref{eq: dso da} with CVXPY for detailed implementation. All the above models are coded by Python 3.10 and deployed in a MacBook Pro with RAM of 16 GB, CPU Intel Core i7 (2.6 GHz). The other detailed data are attached in Ref.~\cite{dataset2023}.

For more precise delivery, we present the organization of the case study with data and experiment setting in section~\ref{sub: data}, calculation efficacy analysis in comparison with RODM in section~\ref{sub: dynamic}, the carbon alleviation effectiveness of the DLME in comparison with CEF in section~\ref{sub: comparison}, the emerging reactive demand response in section~\ref{sub: reactive dr}, and application to an extensive system in section~\ref{sub: large}.

\subsection{Data Processing and Experiment Setting} \label{sub: data}

\subsubsection{Experiment Setting}

We set \emph{i)} the different carbon emission factors for the synchronous machine-based DGs \eqref{eq: sm dg} and the main grid of the 33-bus system \emph{ii)} and other essential SOCP-based scheduling model parameters \eqref{eq: dso da} in Tab.~\ref{tab: opt param}.

\begin{table}[ht]
  \renewcommand{\arraystretch}{1.3}
  \centering
  \caption{The key parameters of DLME calculation in day-ahead scheduling model.}
  \begin{tabular}{cc|cc}
    \hline
    Hyperparameter & Value & Hyperparameter & Value \\
    \hline
    Coal generator $e$ & 0.875 tCO$_2$/MWh & PV capacity & 50MW \\
    Gas generator $e$ & 0.520 tCO$_2$/MWh &  PV number & 5 \\
    Charging depth & 0.5 & Discharging depth & 0.5 \\
    $\eta_{dis}$ & 0.92 & $\eta_{ch}$ & 0.90 \\
    $\Delta_t$ & 1 hour & & \\ 
    \hline
  \end{tabular}
  \label{tab: opt param}
\end{table}
We note that the emission factors from DLME and CEF are normalized into p.u. with the unit of tCO$_2$/MWh.

\subsubsection{Data Processing}

\begin{figure}[ht]
  \centering
  \includegraphics[scale=1]{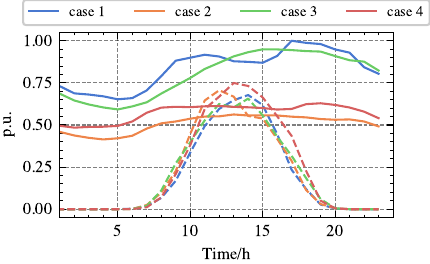}
  \caption{Average load and PV in p.u. under various scenarios with the solid lines for the average load and the dotted lines for the average PV.}
  \label{fig: scenario}
\end{figure}

We leverage the implicit derivation to compute the DLME for two distribution network systems under various renewable and load scenarios. Fig.~\ref{fig: scenario} presents the typical four average load and PV values in the normalized form (p.u.) via the k-mean++ algorithm. 

As shown in Fig.~\ref{fig: scenario}, different colors denote different typical scenarios, the solid lines denote the average load, and the dotted lines denote the average PV. We determine the number of clusters according to the metrics of the sum of the squared Euclidean (SSE) distance.

\subsubsection{Comparison Setting}

To verify the calculation efficacy and carbon alleviation effectiveness, we compare the proposed DLME calculation against \emph{i)} the reduced-order dispatch model (RODM) described in \cite{Deetjen2019} for the distribution network to verify the calculation efficacy of the proposed approach \emph{ii)} and the carbon emission flow (CEF) model proposed in \cite{Kang2015} to verify the effectiveness of the proposed DLME, which are detailed as follows.

The core of the RODM is a merit order-based dispatch process where generators are dispatched in ascending order of cost. After dispatching generators via the merit order, the marginal generator—the generator next in line to modify its output to meet an increase in demand—is identified to find the marginal emissions rate of the system. The corresponding experiment is conducted in section~\ref{sub: dynamic}.

CEF quantifies the power emission responsibility from the power supply to the load demand based on the power flow model formulated in appendix~\ref{apx: cef}. The node carbon intensity (NCI) index in CEF describes the average carbon emission of each bus per unit injected power flow, which can provide the average carbon emission to incentivize low carbon system operation for the system operator. We denote the NCI index of CEF in the distribution network as the distribution locational average emission (DLAE), distinguished from the DLME. The corresponding comparison experiment is conducted in section~\ref{sub: comparison}.

\subsubsection{Budget-based Demand Response Model}

We leverage the budget-based demand response (DR) model to verify the alleviation effectiveness of the proposed DLME under the implicit derivation approach by comparing the carbon reduction performance under various approaches in \eqref{eq: dr}, inspired from \cite{Chen2020}. It \emph{i)} receives the carbon signals from the various models (DLME, RODM, and DLAE), \emph{ii)} maximizes the carbon reduction benefits under the response budget via \eqref{eq: dr budget} for each period, \emph{iii)} and implements the optimized response values for optimal dispatching via \eqref{eq: dr}. 

\begin{eqnarray}
  \allowdisplaybreaks
  \begin{aligned}
    \max \text{ } e^{(\cdot)}_{i} p^{dr}_i \quad \text{s.t. } \sum p^{dr}_i = P^{dr}_{\Delta}
  \end{aligned}
  \label{eq: dr budget}
\end{eqnarray}
where $p^{dr}_{i}$ is the response power of bus $i$; $P^{dr}_{\Delta}$ is total demand response budget; $e^{(\cdot)}_{i}$ is the bus emission signal by various calculation approaches, $(\cdot)$ can be the DLME from implicit derivation and RODM or DLAE from the CEF.

\begin{eqnarray}
  \allowdisplaybreaks
  \begin{aligned}
    & {p}^{D, ca}_{i} = p^{D}_{i} - p^{dr}_{i}
  \end{aligned}
  \label{eq: dr}
\end{eqnarray}
where ${p}^{D, ca}_{i}$ is the power consumption after the demand response, and ${p}^{D, ca}_{i}$ can be taken into the \eqref{eq: dso da} for redispatching and carbon alleviation calculation. Moreover, the demand response capacity of $P^{dr}_{\Delta}$ is set as 1\% of the total system maximum load demand.

\subsection{Calculation Effciency Analysis} \label{sub: dynamic}

This part compares the proposed implicit derivation-based approach with the RODM approach to verify the calculation efficacy of DLME in day-ahead scheduling of \eqref{eq: dso da} from the perspectives of results and response analysis. Concretely, result analysis focuses on the temporal DLME distribution under various PV and load scenarios; response analysis focuses on the carbon alleviation performance through the budget-based DR model under different calculation approaches numerically.

\subsubsection{Results Analysis}

\begin{figure}[ht]
  \centering
  \subfigure[Hourly DLME analysis in case 1.]{\includegraphics[scale=1]{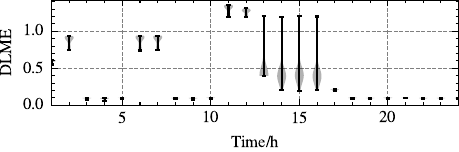}\label{fig: da emission a}}
  \subfigure[Hourly DLME analysis in case 2.]{\includegraphics[scale=1]{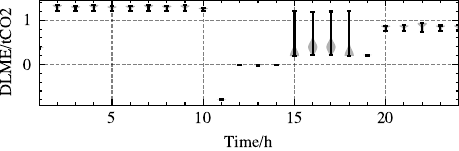}\label{fig: da emission b}}
  \subfigure[Hourly DLME analysis in case 3.]{\includegraphics[scale=1]{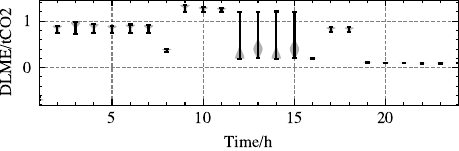}\label{fig: da emission c}}
  \subfigure[Hourly DLME analysis in case 4.]{\includegraphics[scale=1]{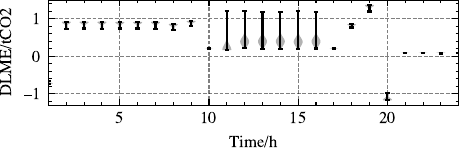}\label{fig: da emission d}}
  \caption{Temporal DLME distribution under different scenarios in violin plot.}
  \label{fig: da emission}
\end{figure}

We leverage the implicit derivation of \eqref{eq: backward} to calculate the DLME for SOCP-based scheduling model \eqref{eq: dso da} under typical four scenarios of Fig.~\ref{fig: scenario} and present the corresponding hourly distribution of DLME via violin plot in Fig.~\ref{fig: da emission}.

As shown in Fig.~\ref{fig: da emission}, the calculated DLME features various spatial and temporal change patterns under various scenarios, demonstrating different carbon reduction potentials.

For calculation efficacy, we compare the DLME from the conventional RODM and the implicit derivation approach in the following Fig.~\ref{fig: rodm comp}. 

\begin{figure}[ht]
  \centering
  \includegraphics[scale=1]{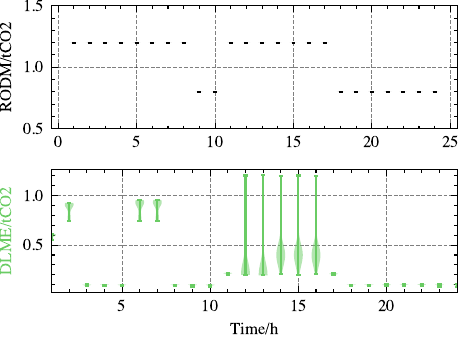}
  \caption{Temporal distribution comparison of DLME from the implicit derivation and RODM in violin plot.}
  \label{fig: rodm comp}
\end{figure}

As shown in Fig.~\ref{fig: rodm comp}, DLMEs under RODM and implicit derivation show a similar pattern with low values during the 9-10 and 21-24 periods. However, DLME under implicit derivation features more temporal and spatial variance to capture the system carbon dynamics, which are more sensitive than the RODM model.

\begin{figure}[ht]
  \centering
  \includegraphics[scale=0.95]{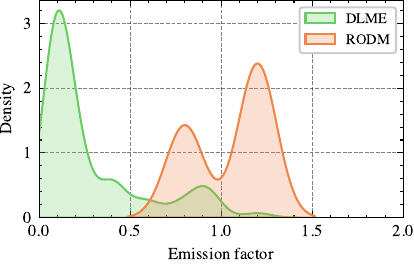}
  \caption{Distribution comparison of DLME from the implicit derivation and RODM in density plot.}
  \label{fig: rodm spatial}
\end{figure}

Fig.~\ref{fig: rodm spatial} further compares the distributions of DLME under the implicit derivation approach and RODM in the kernel density estimate plot, where the emission factors are more concentrated to low values in an implicit derivation way. The carbon alleviation effectiveness of calculated DLME is evaluated via the budget-based DR program.

\subsubsection{Response Analysis}

The above results analysis illustrates the temporal and spatial difference of DLME from the RODM and implicit derivation \emph{intuitively}, where more fluctuation implies high sensitivity. We further verify the carbon reduction effectiveness of DLME from implicit derivation \emph{numerically} based on the budget-based DR model of \eqref{eq: dr budget} and \eqref{eq: dr}. So Tab.~\ref{tab: response da} compares the carbon emission reduction performance via the response model under different DLME calculation approaches.

\begin{table}[ht]
  \renewcommand{\arraystretch}{1.3}
  \centering
  \caption{Comparison of carbon emission reduction performance under various scenarios in the day-ahead scheduling.}
  \begin{tabular}{ccccc}
    \hline
    Cases & Initial/tCO$_2$ & DLME /tCO$_2$ & RODM /tCO$_2$ & Enhance \\
    \hline
    Case1 & 34.006 & 33.660 & 33.815 & 81.525\%
    \\
    Case2 & 17.106 & 16.688 & 16.871 
    & 78.047\%
    \\
    Case3 & 30.987 & 30.563 & 30.781
    & 106.158\%
    \\
    Case4 & 26.099 & 25.852 & 25.906 & 24.761\%
    \\
    \hline
  \end{tabular}
  \vspace{1ex}
  
  {\raggedright We note that ``Enhance'' refers to the improving percentage of carbon alleviation under DLME over other models. \par}
  \label{tab: response da}
\end{table}
As shown in Tab.~\ref{tab: response da}, DLME from implicit derivation achieves lower carbon emission for different cases, verifying its carbon alleviation effectiveness. For carbon alleviation performance, we improve the carbon reduction from RODM with a maximum of 106.16\% and a minimum of 24.76\% enhancement.

\subsection{Comparison of Marginal and Average Emissions} \label{sub: comparison}

This part analyzes the difference between the marginal and average carbon signals by comparing the DLME and DLAE. DLME from implicit derivation \eqref{eq: backward} and DLAE from CEF \eqref{eq: cef node} quantify the carbon emission of the demand side from both the marginal and average perspectives. Similarly, we utilize the result analysis for intuitive analysis and response analysis for carbon alleviation numerical analysis for SOCP-based scheduling of \eqref{eq: dso da}.

\subsubsection{Results Analysis}

We compare the temporal distribution of DLME and DLAE via the violin plot in the following Fig.~\ref{fig: lme ame comp}.
\begin{figure}[ht]
  \centering
  \includegraphics[scale=1]{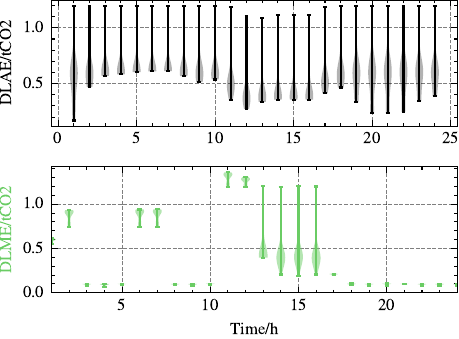}
  \caption{Temporal distribution comparison of DLME and DLAE in violin plot.}
  \label{fig: lme ame comp}
\end{figure}
As shown in Fig.~\ref{fig: lme ame comp}, DLAE fluctuates less than DLME across the scheduling period, indicating it may not capture the carbon reduction dynamics timely. The different patterns imply the different temporal characteristics of DLAE and DLME. The violin plot of DLAE at each period indicates similar values, and, in contrast, the violin plot of DLME at each period indicates the concentrated values from the 1-12 and 17-24 periods and diffused values from the 13-16 periods. The different patterns imply the different spatial characteristics of DLAE and DLME.

Concretely, we compare the temporal change of DLAE and DLME of typical buses (bus 2, 11, 14, and 15) in the following Fig.~\ref{fig: bus comp case 33}. The change of DLAE in these buses is relatively placid, but the DLAE values among these buses are distinguished. In contrast, the change of DLME in these buses is quite varying, but the DLME values among these buses are similar from 1-12 and 17-24 periods. The main reason behind the phenomenon is that DLAE captures the average emission responsibility, and DLME captures the marginal emission responsibility.
\begin{figure}[ht]
  \centering
  \includegraphics[scale=1]{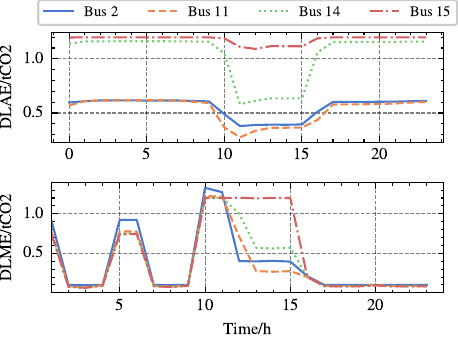}
  \caption{Typical bus temporal comparison of DLAE and DLME.}
  \label{fig: bus comp case 33}
\end{figure}

\begin{figure}[ht]
  \centering
  \includegraphics[scale=0.95]{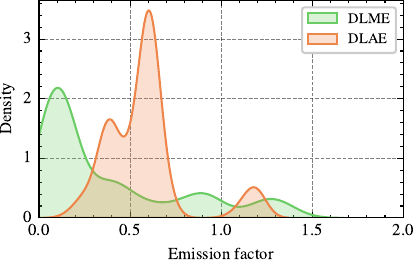}
  \caption{Distribution comparison of DLME and DLAE in density plot.}
  \label{fig: dlae spatial}
\end{figure}

Fig.~\ref{fig: dlae spatial} further compares the distribution of DLAE and DLME in the density plot and indicates the different incentivizing between the average and marginal emission quantification perspectives, whose carbon alleviation effectiveness is verified in Tab.~\ref{tab: me ae}.

\subsubsection{Response Analysis}

We further verify the carbon reduction effectiveness of DLME over DLAE based on the budget-based DR model of \eqref{eq: dr budget} and \eqref{eq: dr} \emph{numerically}. Tab.~\ref{tab: me ae} compares the carbon emission reduction performance under DLAE and DLME via the response model.
\begin{table}[ht]
  \renewcommand{\arraystretch}{1.3}
  \centering
  \caption{Comparison of carbon emission reduction performance under various scenarios in the day-ahead scheduling.}
  \begin{tabular}{ccccc}
    \hline
    Cases & Initial/tCO$_2$ & DLME /tCO$_2$ & DLAE /tCO$_2$ & Enhance \\
    \hline
    Case1 & 32.202 & 31.878 & 31.943 & 25.380\%
    \\
    Case2 & 17.460 & 17.106 & 17.253 & 70.949\%
    \\
    Case3 & 31.326 & 30.987 & 31.052 & 23.946\%
    \\
    Case4 & 26.573 & 26.099 & 26.436 & 244.459\%
    \\
    \hline
  \end{tabular}
  \label{tab: me ae}
\end{table}
As shown in Tab.~\ref{tab: me ae}, DLME from implicit derivation achieves lower carbon emission than DLME for different cases, verifying its carbon alleviation effectiveness. For carbon alleviation performance, we improve the carbon reduction from DLAE with a maximum of 244.46\% and a minimum of 23.95\% enhancement.

For case 1, we further compare the temporal carbon reduction performance comparison under DLME and DLAE in the following Fig.~\ref{fig: cef emission}. Under the same DR budget, different from DLAE, the DLME leverages DR in 5-7 and 14 periods for effective carbon reduction.

\begin{figure}[ht]
  \centering
  \includegraphics[scale=0.9]{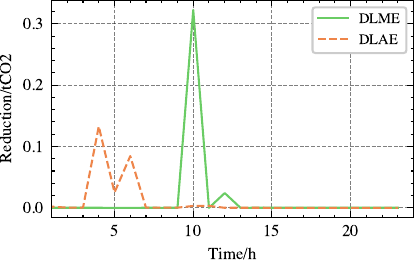}
  \caption{Temporal carbon reduction performance comparison under DLME and DLAE.}
  \label{fig: cef emission}
\end{figure}

\subsubsection{DLME under various RES Penetration} \label{sub: dlme analysis}

With the increase of PV, the penetration of renewables in the distribution is increasing correspondingly. The penetration rate will affect the distribution of the corresponding calculated DLME. So we set the low (20\%), medium (50\%), and high (70\%) PV capacity penetration scenarios for penetration analysis, where the only difference is the PV capacity. The DLME distributions under these scenarios are compared by the density plot in the following Fig.~\ref{fig: dlme penetration}.

\begin{figure}[ht]
  \centering
  \includegraphics[scale=0.95]{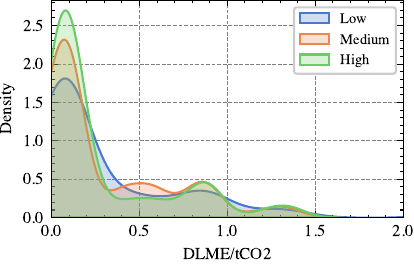}
  \caption{DLME distribution comparison under low, medium, and high renewable penetration in density plot.}
  \label{fig: dlme penetration}
\end{figure}
As shown in Fig.~\ref{fig: dlme penetration},  DLME features the same density distribution pattern but different density values. Higher PV penetration will lead to the concentration of lower DLME values.

\subsection{Reactive DLME for Reactive Demand Response} \label{sub: reactive dr}

Beyond the active power emission, the reactive power demand also generates emission through the active power loss in distributing reactive power of \eqref{eq: dso da}. Future reactive power markets are considered in \cite{PARK2023120303} to unlock the regulation potential of inverter-based DERs to support the system's safe operation. Inspired by \cite{PARK2023120303}, we discuss the carbon reduction potential from the reactive demand response by formulating the reactive DLME in \eqref{eq: DLME reactive}, which is an analogy to active DLME of \eqref{eq: DLME def} and can also be solved via the implicit derivation approach of \eqref{eq: backward}.

\subsubsection{Results Analysis}

\begin{figure}[ht]
  \centering
  \includegraphics[scale=1]{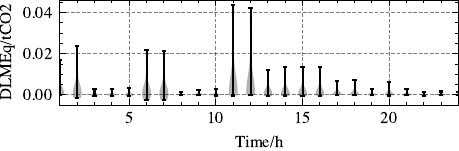}
  \caption{Temporal distribution of reactive DLME in violin plot.}
  \label{fig: dlme reactive}
\end{figure}

We analyze the temporal distribution of reactive DLME via the violin plot in Fig.~\ref{fig: dlme reactive}. 

As shown in Fig.~\ref{fig: dlme reactive}, different from active DLME, the reactive DLME can be negative during the 5-7 periods with different change patterns.

\subsubsection{Reactive Response Analysis}

To verify the reduction effectiveness, we formulate the budget-based reactive DR program in \eqref{eq: dr reactive}, similar to \eqref{eq: dr}.
\begin{eqnarray}
  \allowdisplaybreaks
  \begin{aligned}
    \max \text{ } e^{(\cdot)}_{i} q^{dr}_i \quad
    \text{s.t. } \sum q^{dr}_i = Q^{dr}_{\Delta}
  \end{aligned}
  \label{eq: dr reactive}
\end{eqnarray}

As shown in Tab.~\ref{tab: reactive dr}, compared with active DR, the reactive DR performance for carbon alleviation is less effective with lower emission reduction with maximum 0.00262 tCO$_2$ and minimum 0.00121 tCO$_2$ reductions.

\begin{table}[ht]
  \renewcommand{\arraystretch}{1.3}
  \centering
  \caption{Comparison of carbon emission reduction performance under various scenarios for reactive DLME.}
  \begin{tabular}{ccccc}
    \hline
    Cases & Initial/tCO$_2$ & DLME$_{q}$ /tCO$_2$ & Reduction /tCO$_2$  \\
    \hline
    Case1 & 33.65974 & 33.65711
    & 0.00262 \\
    Case2 & 16.68775 & 16.68650 & 0.00125 \\
    Case3 & 30.56316 & 30.56172 & 0.00144 \\
    Case4 & 25.85825 & 25.85704 & 0.00121 \\
    \hline
  \end{tabular}
  \label{tab: reactive dr}
\end{table}

\subsection{Scalability to Large Systems} \label{sub: large}

\begin{figure}[ht]
  \centering
  \includegraphics[scale=0.95]{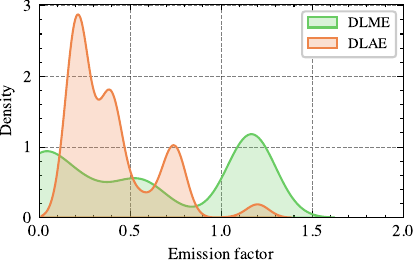}
  \caption{DLME distribution comparison of DLME and DLAE.}
  \label{fig: dlme dlae 69}
\end{figure}

We further verify the scalability of the DLME, and its distribution differs from the DLAE in the modified IEEE 69-bus system, indicating different carbon alleviation incentivizing, as shown in Fig.~\ref{fig: dlme dlae 69}.

Tab.~\ref{tab: me ae 69} further verifies the carbon alleviation enhancement of DLME.
\begin{table}[ht]
  \renewcommand{\arraystretch}{1.3}
  \centering
  \caption{Comparison of carbon emission reduction performance under various scenarios for the 69-bus system.}
  \begin{tabular}{ccccc}
    \hline
    Cases & Initial/tCO$_2$ & DLME /tCO$_2$ & DLAE /tCO$_2$ & Enhance \\
    \hline
    Case1 & 2515.470 & 2341.532 & 2366.841 & 17.028\% \\
    Case2 & 1181.552 & 1035.591 & 1080.098 & 43.870\% \\
    Case3 & 2124.439 & 1978.422 & 1994.626 & 12.483\% \\
    Case4 & 1867.345 & 1780.961 & 1798.437 & 25.361\% \\
    \hline
  \end{tabular}
  \label{tab: me ae 69}
\end{table}
As shown in Tab.~\ref{tab: me ae 69}, DLME achieves a lower carbon emission with a maximum of 43.870\% reduction and a minimum of 12.483\% enhancement, verifying the scalability of DLME for carbon alleviation in the distribution network.

\section{Conclusion} \label{sec: conclude}

In analogy to extending LMPs to DLMPs, this paper establishes the DLME for both active and reactive power to incentivize carbon alleviation in the distribution network. It formulates the SOCP-based day-ahead distribution network scheduling model, analyzes the emission propagation and responsibility from power supply to load demand, and leverages the SOCP-based implicit derivation approach to and calculates the DLMEs effectively in a model-based way. Comprehensive numerical studies are conducted to validate the proposed method, comparing its calculation efficacy to the conventional LME estimation approach, assessing its effectiveness in carbon alleviation compared to the AEFs from CEF, and evaluating its carbon alleviation ability of reactive DLME.

\appendix

\subsection{Carbon Emission Flow Model}\label{apx: cef}

CEF calculation is based on the generation, network, node, and storage side carbon analysis. Based on the Ref. \cite{Kang2015}, the node carbon emission intensity (NCI) with ES is further derived as the weighted average of injected branch flow emission (BCI) and generation (GCI), as illustrated in \eqref{eq: cef node} and \eqref{eq: cef branch}. 

\begin{eqnarray}
  {e}_{i,t} = \frac{\sum_{i\in\mathcal{N}^G_i} {e}^{G}_i {p}^G_{i,t} + \sum_{j\in\mathcal{N}^{l+}_i} {\rho}_{ij,t}| {p}_{ij,t}| + \sum_{i\in\mathcal{N}^{ES}_{i}} p^{dis}_{i,t}e^{dis}_{i,t}}
    {\sum_{i\in\mathcal{N}^G_i} {p}^G_{i,t} + \sum_{j\in\mathcal{N}^{l+}_i} |{p}_{ij,t}| + \sum_{i\in\mathcal{N}^{ES}_{i}} p^{dis}_{i,t}} \label{eq: cef node a}
  \label{eq: cef node}
\end{eqnarray}
where $\mathcal{N}^G_i$, $\mathcal{N}^{l+}_i$, and $\mathcal{N}^{ES}_{i}$ are the sets of generation, flow-in branches, and energy storage with bus $i$; $e_{i,t}$ is the nodal carbon emission intensity for bus $i$ in time $t$.

\begin{eqnarray}
  {\rho}_{ij,t} = \left\{
    \begin{matrix}
      {e}_{i,t} & \quad {P}_{ij,t} \geq 0 \\ 
      {e}_{j,t} & \quad {P}_{ij,t} < 0 
    \end{matrix} 
  \right. \label{eq: cef node b}
  \label{eq: cef branch}
\end{eqnarray}
where $p_{ij,t}$ and $\rho_{ij,t}$ are the active power flow and corresponding BCI from bus $i$ to bus $j$ in time $t$. BCI calculation follows the proportional sharing principle from Ref.~\cite{Kang2015}.

\ifCLASSOPTIONcaptionsoff
  \newpage
\fi

\bibliographystyle{IEEEtran}
\bibliography{IEEEabrv, mybibfile}

\end{document}